\documentclass[aps,pra,twocolumn,showpacs]{revtex4}
\usepackage{amsmath}
\usepackage{amssymb}
\usepackage{epsfig}
\newcommand\Ho{{\hat H}}
\newcommand\ro{{\hat\rho}}

\newcommand\no{{\hat n}}

\newcommand\ah{{\hat a}}
\newcommand\ahd{\ah^\dagger}
\newcommand\Io{{\hat I}}

\newcommand\Lc{\mathcal{L}}
\newcommand\GL{\Gamma_L}
\newcommand\GR{\Gamma_R}
\newcommand{\ket}[1]{\vert #1\rangle}
\newcommand{\bra}[1]{\langle #1\vert}
\newcommand{\half}{{\scriptstyle{\frac{1}{2}}}}

\begin{document}
\title{Simple theory of the measured current through quantum dots}
\author{J\'ozsef Zsolt Bern\'ad}
\email{bernad@complex.elte.hu}
\affiliation{Department of Physics of Complex Systems, E\"otv\"os University, Budapest, Hungary}
\author{Andr\'as Bodor}
\email{bodor@complex.elte.hu}
\affiliation{Department of Physics of Complex Systems, E\"otv\"os University, Budapest, Hungary}
\author{Lajos Di\'osi}
\email{diosi@rmki.kfki.hu}
\affiliation{Research Institute for Particle and Nuclear Physics, Budapest, Hungary}
\author{Tam\'as Geszti}
\affiliation{Department of Physics of Complex Systems, E\"otv\"os University, Budapest, Hungary}
\email{geszti@complex.elte.hu}

\date{\today}

\begin{abstract}
A simple theory of the detected current $I(t)$ flowing through charge qubits ---
quantum dots --- is proposed in terms of standard continuous measurement
theory. Applied to a double dot, our formalism easily confirms previous results on 
quantum Zeno effect, driven by growing ammeter performance $\gamma$.        
Due to the transparent formalism, we can calculate the exact fluctuation spectrum
$S(\omega)$ of the detected current, containing a significant Lorentzian peak near the
Rabi frequency of the double dot. 
\end{abstract}

\pacs{03.65.Ta, 03.65.Xp, 73.23.-b}

\maketitle

Back-action of measurement on the measured object is an emblematic property of quantum systems.
Quantum Zeno effect (QZE) - suppression of the object's coherent internal dynamics by measurement -
is one of the marked scenarios for that back-action. In this Letter we are going to discuss the
effect in the framework of unsharp measurements, for which there is a well-developed theoretical 
framework called time-continuous weak measurement theory (cf. Refs.~\cite{Dio06,JacSte07} and 
references therein). The application we have in mind is the case of a double quantum dot (DQD),
which is a semiconducting nanostructure, available in high quality due to massive progress in 
experimental technology. 

There is a growing culture of indirect measurements on nanostructures by means of Coulomb-coupled 
quantum point contacts, single-electron transistors, or DQD's \cite{Gur97}-\cite{Mao04}. QZE is one 
of the effects studied from the beginning \cite{Gur97} 
(cf. also \cite{QZE}), and the concept of time-continuous measurement has penetrated the field 
\cite{Gur97,Kor99,Goa01} for a long time. All those studies assume the -- sharp or unsharp -- 
detection of the number $N(t)$ of electrons that have tunneled through the nanostructure, and the 
current $I(t)$ is defined as the stochastic mean $\langle dN(t)/dt\rangle$. The present work differs 
from those studies in  assuming that detection is done by  a tool usually considered as fully classical: 
an ammeter of high performance, monitoring the time-dependent current $I(t)$ flowing through that device.

The main parameter of the theory is detection performance (or detection strength), defined as
\begin{equation}
\label{gamma}
\gamma=(\Delta t)^{-1}(\Delta I)^{-2}~, 
\end{equation}
where $\Delta t$ is the time-resolution (or, equivalently, the inverse bandwidth) of the ammeter and 
$\Delta I$ is the statistical error characterizing unsharp detection of the average current in the period 
$\Delta t$. The accuracies of commercial ammeters reach  $pA\approx 10^7~electron/s$, at a bandwidth 
of $10^4$ Hz. Then $\gamma=10^{-10}~s$, which should be compared to the time scale of internal coherent 
dynamics of a DQD, characterized by the Rabi frequency $\Omega \approx 10^{10}$ Hz, which is also the order 
of magnitude of the steady-state current measured in $electron/s$ units, \emph{viz.} $I/e$. 
After all, $\gamma\Omega\approx 1$ can be reached through standard instrumentation. Below we are going to 
show that this suffices for observing a continuous-measurement version of QZE. 

We rely upon the standard Markovian approximation, tracing out environmental variables,
and treating transport in terms of the reduced density matrix $\ro$ of the DQD and a corresponding effective 
current operator $\Io$. Then the stochastic mean of the detected current is obtained as the quantum 
mechanical subsystem average \cite{foo1}
\begin{equation}\label{currentmeas}
\langle I(t)\rangle=\langle\Io\rangle_{\ro(t)}.
\end{equation}
Continuous measurement theory starts from this point, and accounts for a double action of the ammeter of 
finite performance:

1. The quantum mechanical back-action on the continuously measured quantum system, our main concern here, 
induces loss of coherence between eigenstates belonging to different eigenvalues of the measured quantity; 
the rate of decoherence being proportional to the detection performance $\gamma$. Under well-defined 
conditions \cite{Dio06,JacSte07} it can be represented by adding a Lindblad-type 
term to the master equation:
\begin{equation}\label{mastermeas}
\frac{d\ro}{dt}=\Lc\ro-\frac{\gamma}{8}[\Io,[\Io,\ro]]\equiv\Lc'\ro,
\end{equation}
Our analysis is based on the solution of this extended master equation, in which the Lindblad 
supermatrix $\Lc$ provides a Markovian description of decoherence and damping, as usual. 

2. Along with the quantum effect we expect to observe, inaccurate measurement generates fully classical 
white noise in the measured current, superposed on the average 
(\ref{currentmeas}): 
\begin{equation}\label{currentnoise}
I(t)-\langle I(t)\rangle=gw(t).
\end{equation}
Here $w(t)$ is white noise on the scale of the Markovian dynamics (i.e., a flat spectrum
noise of band-width much larger than the typical frequency range correctly accounted for by 
the Markovian approximation). The amplitude of the noise is related to the detection performance $\gamma$
according to an inverse square-root law:
\begin{equation}\label{gdef} 
g=1/\sqrt{\gamma}.
\end{equation}

Before doing the calculation, we observe that a genuine ammeter, unlike an electron counter, smears out 
dynamical details of shot noise originating from incoherent tunneling of individual electrons through the 
external borders of the DQD, while retaining time structure due to internal coherent dynamics (Rabi 
oscillations). Accordingly, we omit shot noise from the present treatment; hence the calculus becomes 
surprizingly simple and robust, as can be easily checked for a single dot where the power spectrum is exactly 
known from the $N$-resolved calculations \cite{bjzsba}. 

Let us introduce the stationary solution $\ro_\infty$ of the extended master equation (\ref{mastermeas}):
\begin{equation}\label{defrhostac}
\Lc'\ro_\infty=0~,
\end{equation} 
as well as the stationary current:
\begin{equation}\label{defIstac}
I_\infty=\langle I(t)\rangle_\infty=\langle \Io\rangle_{\ro_\infty}~,
\end{equation} 
where $\langle.\rangle_\infty$ stands, in general, for stochastic mean of currents detected on 
$\ro_\infty$.
We also define the Heisenberg operator of the current for $t>0$:
\begin{equation}\label{currentHeis}
\frac{d\Io(t)}{dt}=(\Lc')^\dagger\Io(t)~,~~~~~~\Io(0)=\Io~.
\end{equation}
In continuous measurement theory, the stationary correlation function of the fluctuating
detected current turns out to be: 
\begin{equation}\begin{split}\label{corr}
&\langle I(t)I(0)\rangle_\infty - \langle I(t)\rangle_\infty\langle I(0)\rangle_\infty \\
&~~~~~=\mathrm{Re}\langle\Io(\vert t\vert )\Io\rangle_{\ro_\infty}-I_\infty^2~
+g^2\delta(t)~.
\end{split}\end{equation}
The nonsingular term on the r.h.s. follows from the standard expression 
$\langle I(t)I(0)\rangle_\infty = \mathrm{Re}\langle\Io(\vert t\vert )\Io\rangle_{\ro_\infty}$
valid for bulk quantum systems where the detector noise can be neglected. For our nanostructure,
the singular term at $t=0$ comes from the white-noise fluctuations, originating from unsharp measurement, 
i.e., detection noise, introduced in eq. (\ref{currentnoise}). We have thus expressed the stationary 
correlation function of the detected classical current $I(t)$ in terms of the quantum correlation of 
the Heisenberg current $\Io(t)$. The spectral density of the detected fluctuations will be defined as the 
Fourier transform \cite{foo2} of the correlations (\ref{corr}), resulting in the form:
\begin{equation}\label{fluctspectr}
S(\omega)=\int\langle I(t)I(0)\rangle_\infty {e}^{i\omega t}dt - 2\pi I_\infty^2\delta(\omega)~,
\end{equation}
which for high frequencies approaches the measurement-added white noise value 
\begin{equation}\label{sinfty}
S(\infty)=g^2=1/\gamma.
\end{equation}
For \emph{very high} values of ammeter performance $\gamma$ that noise level may get below the ``shot noise 
limit'' $I_{\infty}/2$. In that case ammeter and charge counter measurements may differ in characteristic 
ways not controlled by the present method of calculation; however, this is not our concern here, since the 
effect we envisage remains in the well-treated range of moderately high $\gamma$.

Now we turn to the specific features of the double quantum dot, consisting of two potential wells 
(``dots'': left and right), with an internal barrier allowing coherent tunneling between them, and two 
external barriers allowing incoherent tunneling between each dot and its joining lead. The Markovian 
approximation used is based on the assumption thermalization in leads L and R being the fastest dynamical
process present.

Because of intra-dot Coulomb blocade, at low temperature in each of the dots there can be 
but one electron in the ground state, or none. Strong inter-dot Coulomb repulsion further reduces the set of 
available orthogonal basis states to the following three:
\begin{equation}
\ket{0}\equiv\ket{0,0},~~~ \ket{L}\equiv\ket{1,0},~~~\ket{R}\equiv\ket{0,1}.
\end{equation}
On the above basis, we introduce the following absorption and emission operators:
\begin{equation}
\ah_L=\ket{0}\bra{L},~~~\ah_R=\ket{0}\bra{R}~,
\end{equation}
as well as the charge operators:
\begin{equation}
\no_L=\ket{L}\bra{L},~~~\no_R=\ket{R}\bra{R},~~~\no=\no_L+\no_R=1-\ket{0}\bra{0}~.
\end{equation}
With the above definitions, the Hamiltonian is this:
\begin{equation}\label{H0}
\Ho_0=\half\delta(\no_L-\no_R)+\Omega(\ahd_L\ah_R+H.C.)~.
\end{equation}

Neither the Hamiltonian, nor any of the relevant observables (in particular, the current -- see below) have 
off-diagonal elements connecting the state $\ket{0}$ to the rest of the reduced Hilbert space of the DQD, 
imposing an effective charge superselection. Therefore the off-diagonal elements $0L,0R,L0,R0$ of the matrix 
$\ro$ should be set to zero identically:
\begin{equation}\label{rhodd}
\ro_\infty=\left(\begin{array}{ccc}\rho_{00}&0&0\\0&\rho_{LL}&\rho_{LR}\\0&\rho_{RL}&\rho_{RR}
\end{array}\right)~.
\end{equation}

The Markovian approximation yields two irreversible processes that modify the intrinsic Hamiltonian dynamics 
of the dots: tunneling of an electron from lead L to the left dot at rate $\GL$, and tunneling of an electron 
from the right dot to lead R at rate $\GR$. The rates $\GL,\GR$ depend on the details of the total Hamiltonian 
dynamics of L+R+dots, and we shall take their value for granted. The two processes contribute to two currents, 
$\Io_L=\GL(1-\no)~$ and $\Io_R=\GR\no_R$. Taking into account the Ramo-Shockley effect of fast screening in 
the leads \cite{RS}, on the slow Markovian time scale there is a single time-local 
operator of the current flowing from  L through dots and R to the ammeter:
\begin{equation}\label{current}
\Io=\frac{\GL(1-\no)+\GR\no_R}{2}
~=~\frac{1}{2}\left(\begin{array}{ccc}\GL&0&0\\0&0&0\\0&0&\GR\end{array}\right)~.
\end{equation}
The Lindblad supermatrix $\Lc$ appearing in eq.~(\ref{mastermeas}) contains the modification of  
Hamiltonian dynamics due to the two tunneling currents $\Io_{L/R}$. It is easy to work out:
\begin{eqnarray}\label{masteqDD}
&&\Lc\ro\equiv-i[\Ho_0,\ro]+\\
&&+\GL(\ahd_L\ro\ah_L-\half\{\ah_L\ahd_L,\ro\})+\GR(\ah_R\ro\ahd_R-\half\{\ahd_R\ah_R,\ro\})\nonumber
\end{eqnarray}
where Lamb shifts have been included in the energy split $\delta$ of $\Ho_0$. Note that the continuity 
equation is always satisfied \cite{GebCar04,BodDio06}:
\begin{equation}\label{continuity}
\Io_R-\Io_L+\Lc^\dagger\no=0~,
\end{equation}
where $\Lc^\dagger$ is the adjoint of $\Lc$.

Quantum measurement back-action enters through the modified Lindblad supermatrix $\Lc'$ introduced in  
eq.~(\ref{mastermeas}).  Then the stationary state is obtained by solving eq.~(\ref{defrhostac}),
in which eqs. (\ref{mastermeas}), (\ref{H0}), (\ref{current}) and (\ref{masteqDD}) are used as input.
The calculation is standard; surprisingly, we obtain exactly the same equations as those derived by 
Gurvitz \cite{Gur97} for the closely related but still different model of a DQD observed by a  
point-contact charge detector, if Gurvitz's clicking rate $\Gamma_d$ is replaced by
our combination $\gamma \GR^2/16$. We think the reason
is that as long as Markovian approximation is valid, the effective current operator $\Io$ is pinned down to the 
single-dot occupation operators $\no_{L/R}$ through eq.~(\ref{current}). Then a point-contact charge counting device,
like the one discussed by Gurvitz, and a commercial ammeter, as we suggest, have no more freedom than to couple
to the DQD in the same way.

The stationary electron number current reads:
\begin{equation}\label{Istac}
I_\infty=\frac{\GL\GR\Omega^2(1+y)}
{\delta^2\GL+(2\GL+\GR)\Omega^2(1+y)+\frac{1}{4}\GL\GR^2(1+y)^2 },
\end{equation}
where $y=\gamma\GR/16$. The above formula contains the way the Quantum Zeno Effect appears with growing measurement 
performance $\gamma$ (see Fig. 1): in all cases, $I_\infty\to 0$ for $\gamma\to\infty$. By inspection of the density
matrix we learn that this is due to damping of the coherent interdot transport, causing increased occupation of the 
left dot, blocking the way of new electrons to enter. With no bias, $\delta=0$, 
the reduction of current is monotonous; with sufficiently strong bias, $\delta/\Omega>2$, however, there is a 
range of small performances where the current \emph{increases} with growing $\gamma$ (``anti-Zeno effect''). 
Apparently, asymmetric occupation induced by decohering measurement and that caused by bias are competing with 
each other. The whole effect is small though: for $\delta=0$, as the measurement performance passes the shot-noise 
limit $\gamma\approx2/I_\infty$, it reaches a  $3\%$ reduction of current.

\begin{figure}
\epsfig{file=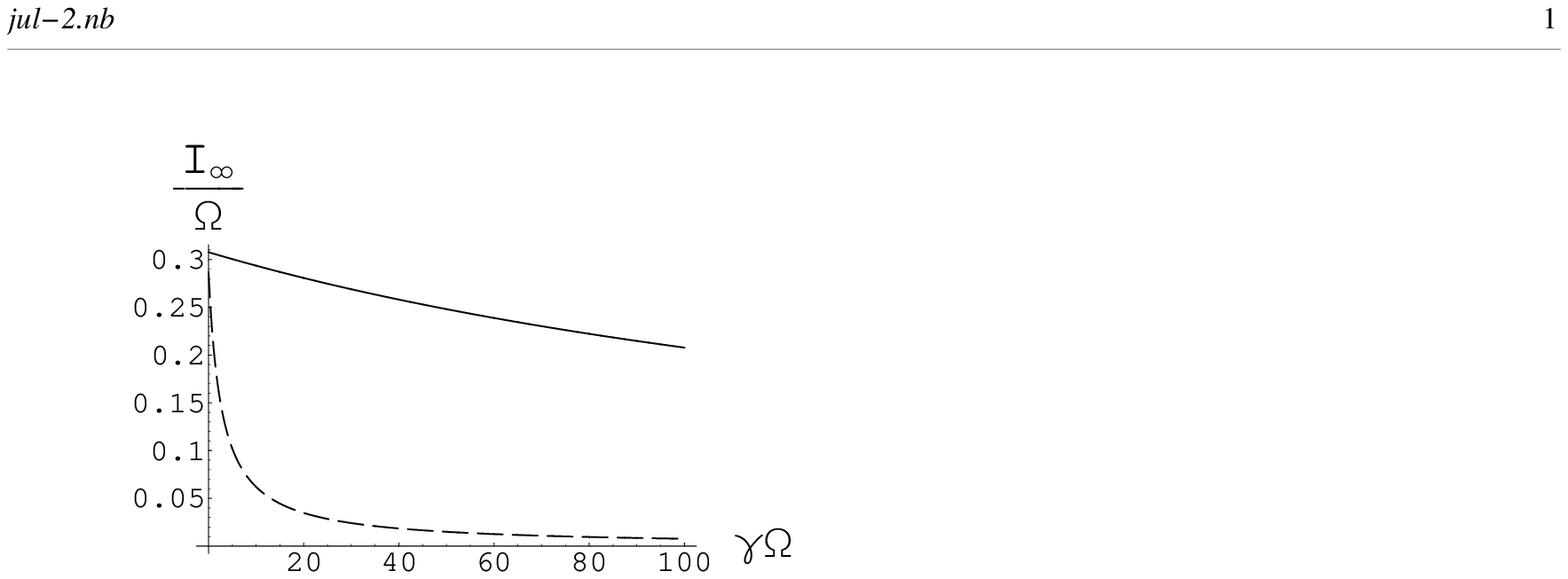=width\hsize,clip=,bbllx=100,bblly=550,
bburx=500,bbury=720}
\caption{\label{FIG.1} Measured stationary current through an unbiased double quantum dot, 
as a function of ammeter performance, for symmetric tunneling rates $\Gamma_L=\Gamma_R=\Omega$ 
(eq. (\ref{gamma})) (solid line), and asymmetric ones with $\Gamma_R=5\Gamma_L=\Omega$ (dashed line). 
The shot-noise limit is at $\gamma\Omega=6.7$.
} 
\end{figure}

Of particular interest for the experiment is the case of asymmetric tunneling rates, $\Gamma_R>\Gamma_L$, 
resulting in much stronger QZE, reaching 30\% for the example displayed in the figure. In this case the right 
dot is depleted with respect to the left one, which causes higher sensitivity to measurement-induced blocking 
of the left-to-right tunneling. Asymmetry of the opposite sign has no marked effect on QZE.

Our simple theory presents the best of its performance in evaluating unsharp measurement back-action 
effects on the detected current fluctuation spectrum. Equations (\ref{corr}) and (\ref{fluctspectr}) set 
the framework for the calculation; in particular, it is the operator correlation function appearing in 
eq.~(\ref{corr}) that has to be evaluated by means of our Markovian master equation. We outline the 
unbiased case $\delta=0$; then 
\begin{equation}
\Lc'(\ket{L}\bra{R}+\ket{R}\bra{L})=-\half\GR(1+\gamma\GR/16)(\ket{L}\bra{R}+\ket{R}\bra{L})
\end{equation}
hence the real part of the $LR$ component of the Heisenberg current $\Io(t)$ defined by 
eq.~(\ref{currentHeis}) remains zero. So the matrix $\Io(t)$ has four independent components instead of five. 
We can thus replace eq.~(\ref{currentHeis}) by $d\Io(t)/dt={\Lc'_I}^\dagger\Io(t)$ where
$\Lc'_I$ is the restriction of $\Lc'$ on the four-dimensional subspace explored by $\Io(t)$.
Note, furthermore, that $\Lc'$ is degenerate also because $\Lc'\ro_\infty=0$ and this degeneracy is inherited
by $\Lc'_I$, too. Therefore the characteristic equation $\vert\Lc'_I+\lambda\vert=0$ 
has a trivial zero root, and the rest of the characteristic equation is only cubic. For the
special case $\GL=\GR=\Omega$ it turns out to be:
\begin{equation}
\lambda^3 -\left(\!\frac{5}{2}+\frac{\gamma \Omega}{32}\right)\!\Omega\lambda^2
          +\left(\!\!6+\frac{\gamma \Omega}{16}\!\right)\Omega^2\lambda
          -\left(\!\frac{13}{2} + \frac{\gamma \Omega}{32}\!\right)\!\Omega^3\!=\!0~.
\end{equation}
It has one real positive root $\lambda=\Delta_0$ and two complex conjugate roots 
$\lambda=\Delta_1\pm i\omega_R$ with positive $\Delta_1$, implying that the solution $\Io(t)$ 
consists of an exponentially decaying and an exponentially damped oscillatory part, the latter
oscillating at the damped Rabi frequency $\omega_R\approx 2\Omega$. 
Let us have a look at the structure of the corresponding fluctuation spectrum (\ref{fluctspectr}):
\begin{equation}
\label{Somega}
S(\omega)\!\!=\!\!\frac{R_0}{\omega^2+\Delta_0^2}
              \!+\!\frac{R_1+\omega R_1'}{(\omega-\omega_R)^2+\Delta_1^2}
              \!+\!\frac{R_1-\omega R_1'}{(\omega+\omega_R)^2+\Delta_1^2}+g^2~.
\end{equation}
One can obtain closed expressions for the parameters of the above expression. We present the results 
for low values of the dimensionless combination $\gamma\Omega$:
\begin{eqnarray}
\omega_R/2\Omega&&=1.025+0.002 \gamma\Omega + {\cal O}(\gamma\Omega)^2, \nonumber \\
\Delta_1/\Omega&&=0.52+ 0.014 \gamma\Omega + {\cal O}(\gamma\Omega)^2.
\end{eqnarray}
This spectrum has one local minimum at $\omega=0$ and two wide peaks at the Rabi 
frequency $\pm\omega_R$, the widths of the peaks are cca. one fourth of the Rabi frequency, 
and on increasing the performance of the ammeter their amplitudes start to decrease. For high
values of $\gamma$ the Rabi peaks get overdamped and merge into a single peak around  $\omega=0$,
see Fig. 2, where background noise has not been included. We stress that to observe the peaks themselves
a good ammeter (high $\gamma$) is needed, since a poor ammeter introduces too much background 
white-noise $g^2=1/\gamma$ to see anything of the coherent peak structure.

\begin{figure}
\epsfig{file=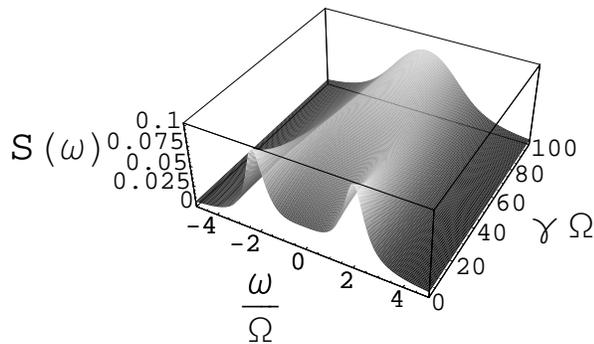=width\hsize,clip=,bbllx=80,bblly=550,
bburx=500,bbury=720}
\caption{\label{FIG.2} Noise spectrum of the measured double-dot current, as a function of the
ammeter performance $\gamma$.} 
\end{figure}

As a conclusion, we have derived explicit expressions for the back-action of an ammeter
on measurable characteristics of a double quantum dot. The back-action has the character of quantum Zeno effect, 
counteracting coherent internal motion of the object, revealed both as reducing the mean transmitted current, 
and damping - eventually, overdamping - Rabi oscillations, as observed in the noise spectrum. In the system 
studied, measurement back-action takes place through decoherence-controlled asymmetry in the occupation 
probabilities of the two coupled quantum dots. Other effects influencing the same asymmetry may compete with 
the effect, eventually producing ranges of anti-Zeno effect. Appropriately tuned asymmetric tunneling barriers 
may strongly enhance the possibilities of observing the effect.


This work was supported by the Hungarian OTKA Grant No. 49384. The authors thank Andr\'as Halbritter 
for useful discussions.


\begin{thebibliography}{99}
\bibitem{foo2} Note that $S(\omega)$ is sometimes defined with an additional factor $2$.
\bibitem{foo1} In the present formalism it takes some extra care to distinguish $\langle.\rangle$ 
that means the average of a random number, from $\langle\hat.\rangle$ (with a hat) that denotes the 
quantum mechanical average of an operator.
\end{thebibliography}
\end{document}